\documentclass[11pt,twoside]{article}
\usepackage{graphicx,epsfig,natbib,epstopdf}
\usepackage{CS18}

\newcommand{\be}{\begin{equation}}
\newcommand{\ee}{\end{equation}} 
\def\lta{\,\raise 0.3 ex\hbox{$ < $}\kern -0.75 em
 \lower 0.7 ex\hbox{$\sim$}\,}
\def\gta{\,\raise 0.3 ex\hbox{$ > $}\kern -0.75 em
 \lower 0.7 ex\hbox{$\sim$}\,}  

\begin{document}

\title{Magnetically Controlled Mass Loss from Exoplanets}

\author{Fred C. Adams$^{1,2}$ and James E. Owen$^{3}$} 

\affil{$^1$Michigan Center for Theoretical Physics \\
Physics Department, University of Michigan, Ann Arbor, MI 48109} 

\affil{$^2$Astronomy Department, University of Michigan, Ann Arbor, MI 48109} 

\affil{$^3$Canadian Institute for Theoretical Astrophysics \\
University of Toronto, 60 St. George Street, Toronto, Ontario, Canada} 

\begin{abstract}
Hot Jupiters can experience mass loss driven by heating from UV
radiation from their host stars, and this flow is often controlled by
magnetic fields. More specifically, near the planetry surface, the
magnetic pressure dominates the ram pressure of the outflow by several
orders of magnitude. After leaving the vicinity of the planet, the
flow must connect onto the background environment provided by the
stellar wind and the stellar magnetic field.  This contribution
considers magnetically controlled planetary outflows and extends
previous work by comparing two different geometries for the background
magnetic field provided by the star. In the first case, stellar field
is assumed to retain the form of a dipole, which is anti-aligned with
the dipole field of the planet.  In the second case, the stellar
outflow opens up the stellar magnetic field structure so that the
background field at the location of the planet is perpendicular to the
planetary dipole. Using numerical simulations, we consider the launch
of the planetary wind with these field configurations.
\end{abstract}

\section{Introduction}

Thousands of alien worlds have now been discovered. As the galaxy-wide
planetary census continues to grow, we can probe their physical
properties, orbital dynamics, chemical composition, and even their
weather. Although they are not the most common planets, Hot Jupiters
with orbital periods of 2 -- 5 days play a vital role: Almost
everything that is currently known about the physical --- as opposed
to orbital --- properties of extrasolar planets has been obtained from
the subset of these planets that are observed to transit bright stars.

The galactic inventory of Hot Jupiters has increased steadily. Nearly
600 of the exoplanets discovered to date orbit their stars with
periods of 10 days or less.  Conventional wisdom \citep{lisssteve}
holds that these planets formed at larger distances from their parent
stars and subsequently moved inward through the action of disk
migration \citep{linbodrich,papaterq}, planet-planet scattering
\citep{rasioford,moorhead}, or Kozai cycles with tidal friction
\citep{eggleton,wumurray,fabrycky}. After migration, the planets
become stranded at small semimajor axes ($a\lta0.1$ AU). 

After the planets reach their inner orbits, they are subjected to
intense heating from the central star.  This heating, which is most
effective for UV photons, can drive photo-evaporative flows from the
planetary surfaces and also generates ionization levels high enough so
that MHD effects play a role. In extreme cases, planetary mass loss
can affect both the final masses and densities of the planets. A good
example is provided by the observed Roche lobe overflow from WASP-12b.
Planets of lower mass are influenced to a greater degree
\citep{owenwu}. Moreover, outflows can be observable even if their
effect on the final mass is modest, and can provide important
information about planetary properties.  Mass loss from extrasolar
planets has already been observed for two transiting planets
associated with bright stars: HD209458b \citep{vidal} and HD189733b
\citep{estangs2010}.

Theoretical calculations of mass loss from extrasolar planets have
shown a steady progression. Pioneering models of outflows from these
planetary bodies have been constructed \citep{lammer,barf1,barf2} and
indicate that substantial mass loss can take place.  These studies
primarily use energy-limited outflow models \citep{watson}, in
conjunction with physically motivated scaling laws, and predict a
range of outflow rates for given planetary masses and external UV
fluxes. The next generation of calculations considered refined
treatments of the chemistry, photoionization, and recombination
\citep{yelle,garcia} as well as including the effects of tidal
enhancement \citep{mc2009}. Next, two-dimensional effects (allowing
non-spherical geometry) in planetary winds were considered
\citep{stoneproga} and indicate that the mass loss rates can be less
than those in the spherical limit. Alternative explanations of the
observations have also been put forth, wherein the inferred excess
material is due to a confined exosphere \citep{trammell2011} or a mass
transfer stream \citep{lai}. However, the observations show high
velocity signatures, which indicate winds or outflows, rather than
static structures (see also \citealt{holmstrom}). For completeness, 
we also note that interactions between close planets and stellar 
magnetospheres can produce interesting observable signatures 
\citep{shkolnik2005,shkolnik2008,lanza08,lanza09}.

Magnetic fields with moderate strength ($B\gta0.3$ gauss) are
sufficient to completely dominate the ram pressure of the winds and
thereby control the flow. In spite of this dominance, however,
magnetic fields are generally ignored in models of planetary winds.
The first studies including magnetic fields considered analytic and
semi-analytic treatements of flow \citep{adams2011,trammell2011}, and
found that not all field lines are active; this reduction leads to
suppression of the mass loss rate. Subsequent numerical studies
confirm this finding \citep{trammell2014,owenadams}, and indicate that
the the flow is magnetically controlled even for the highest expected
UV fluxes (with moderate field strengths) and that outflow from the
night side of the planet is also suppressed. The aforementioned
numerical studies consider the magnetic field of the background star
to be anti-aligned with the planetary dipole \citep{owenadams} or do
not include it \citep{trammell2014}. Nonetheless, the background field
geometry can play an important role. This contribution compares two 
choices for the background mangetic field provided by the star. In the
first case, we consider the field to be anti-aligned (as in
\citealt{owenadams}). In the second case we assume that the stellar
wind is strong enough to open up the stellar magnetic field into a
split-monopole configuration; as a result, the background field at the
location of the planet is perpendicular to the planetary dipole.

\section{Basics} 

We start by outlining the basic physics of the problem: If the UV flux
from the star provides sufficient heating, so that the sound speed at
the planetary surface is greater than the depth of the potential well,
$a_S^2 > G M_P / R_P$, then mass loss occurs without suppression. The
outflow rate is then roughly given by ${\dot M}=4\pi{R_P^2}a_S\rho$,
where $\rho$ is the gas density at the radius $R_P$. In practice,
however, the sound speed is smaller than this benchmark value.  For
typical systems, the sound speed $a_S \approx 10$ km/s and the escape
speed is $\sim50-60$ km/s, so that $a_S^2<GM_P/R_P$ and outflow can
occur only at larger radii, above the planetary ``surface'' at $R_P$.
At these larger radii, the gas temperatures are higher because the
stellar UV flux is less attenuated, and the gravitational potential 
well is shallower. Both of these trends act to increase mass loss. 
However, the gas density decreases rapidly with radius (above the
planetary surface) and this lower density leads to a lower mass loss
rate. As a result, outflow rates $\dot M$ are determined by a delicate
balance between the UV heating rate and the depth of the gravitational
potential well at the launching point.  To leading order, we can
estimate the expected outflow rate as follows: If the outflow is
limited by the rate at which the gas gains energy from the stellar UV
flux, the mechanical luminosity of the outflow must balance the rate
of energy deposition, 
\be 
{GM_P{\dot M} \over R_P} = \varepsilon F_{UV}\pi R^2\,, 
\ee 
where $\varepsilon$ is an efficiency factor and $R=\alpha{R_P}$
determines the area over which energy is absorbed. The mass outflow
rate is thus given by 
\be 
{\dot M} = \varepsilon \alpha^2 \pi {R_P^3 F_{UV} \over (G M_P)}\,.
\label{estimate} 
\ee
For a Jovian planet in a 4-day orbit about a Sun-like star, we expect
$F_{UV}\approx450$ erg cm$^{-2}$ s$^{-1}$, and the expected mass
outflow rate ${\dot M} \sim 10^{10}$ g/s. This value is in rough
agreement with the inferred outflow rates for the planets HD209458b
and HD189733b, indicating that planetary mass loss driven by UV
heating is plausible. However, the outflow process is more complicated
than this idealized picture.

The defining feature of this work is the inclusion of magnetic fields,
which are strong enough to influence the flow
\citep{adams2011,trammell2011}. These fields arise from both the star
and the planet.  When the outflow follows the magnetic field lines,
the flow geometry is set by the field structure, which can be quite
complicated. In particular, the outflows depart significantly from
spherical symmetry and previous (primarily spherical) wind models are
not applicable.  In spite of this complication, the outflow problem
can be reduced to one flow dimension by constructing a new coordinate
system where one coordinate follows the magnetic field lines. This
approach allows for the outflow properties and the passage through the
sonic points to be determined semi-analytically \citep{adams2011,ag2012}. 

To leading order, the stellar and the planetary magnetic fields are
expected to have dipole forms. Typical field strengths on the surfaces
of stars that host Hot Jupiters are measured to be $B_\ast\sim40$ G
(e.g., for HD189733; \citealt{fares}). The surface field strength for
the planets are expected to be $B_P \sim 10$ G (with a factor of
$\sim10$ variation; see \citealt{batygin} and references therein).
Near the planet, we expect that its dipole field dominates, and the
stellar field provides a background field of nearly constant strength
and direction (as seen from the scale of the planet). The simplest
case where the background field of the star is anti-aligned with
planetary dipole can be addressed analytically. An important
complication arises because the dipole field of the star can be opened
up by the stellar wind, so that the background field (from the star)
becomes nearly radial (split monopole configuration). This effect can
be modeled via potential fields with multiple components
\citep{adams2011,ag2012}, or by introducing a source surface
\citep{gregory2010,gregory2011}.  In either case, the stellar field
transitions from a dipole to a split monopole configuration at a
(relatively) well-defined radius $r_S$. Observations of star-planet
interactions are starting to put constraints on field geometries, but
planetary orbits (with 4 day periods and $a \sim 0.05$ AU) could lie
on either side of the transition radius $r_S$ \citep{fares}. In
addition, if the dipole is highly tilted with respect to the orbital
plane, stellar field can be essentially radial at the position of the
planet, even if the field lines are not opened up. As a result,
planetary outflows are expected to experience a wide range of
possibilities for the background magnetic configurations provided by
their host stars.

In order for the magnetic field to influence the outflow, the plasma
must be well-coupled to the field, which requires the cyclotron
frequency $\omega_C$ to be larger than the collision frequency
$\Gamma_C$.  Straightforward calculations indicate that $\omega_C \gg
\Gamma_C$ from the planetary surface all the way out to $r \sim 10^4
R_P$, so that the outflow will indeed be well-coupled to the magnetic
field \citep{adams2011,trammell2011}. Another necessary condition for
the magnetic field to guide the outflow is that the magnetic pressure
must be larger than the ram pressure of the flow, i.e., $B^2/8\pi \gg
\rho v^2$. The ram pressure is given by the outflow rate (${\dot M}
\sim 10^{10}$ g/s) in conjunction with the flow speed ($v \sim a_S
\sim 10$ km/s).  The magnetic field can be approximated as a dipole,
with surface field strength comparable to Jupiter ($B_P \approx 4$ G)
and the usual spatial dependence $B \propto r^{-3}$. For typical
parameters, we find that the magnetic pressure is larger than the ram
pressure of the outflow by a factor of $\sim10^4$ at the sonic surface
and by a factor of $\sim10^6$ at the planetary surface.  These
considerations show that the magnetic field is well-coupled and that
{\it the outflow must be magnetically controlled}.

For this work, we consider the magnetic field of the planet to be a
dipole oriented in the $\hat z$ direction. The outflow takes place on
the spatial scale of the planetary radius $R_P$, which is much smaller
than the stellar radius $R_\ast$ and the orbital scale $a$. As a
result, for purposes of studying the launch of the outflow, the
background magnetic field of the star can be considered as a uniform
field with a constant direction. The total field thus can be written 
\be
{\bf B} = B_P \left[ \xi^{-3} 
\left( \cos\theta {\hat r} - {\hat z} \right) 
+ \beta {\hat s} \right]\,,
\label{bfield} 
\ee
where $B_P$ is the surface field strength of the planet and we have
defined $\xi\equiv{r}/R_P$. The parameter $\beta$ sets the strength of
the background stellar field (evaluated at the orbit of the planet)
and $\hat s$ defines its direction (here we take either ${\hat
  s}={\hat z}$ or ${\hat s}={\hat x}$). Since both the star and planet
are expected to have surface field strengths of a few gauss, and since
Hot Jupiters orbit at distances $a\sim10 R_\ast$, we expect the
parameter $\beta\sim10^{-3}$.

\section{Numerical Calculations}

In order to study magnetically controlled outflow from planets, as
outlined in the previous section, we perform simulations of the
outflow problem \citep{owenadams}. In this numerical treatment, we
solve the Radiation-MHD problem in the ideal MHD limit, i.e., the
magnetic structure is allowed to respond to the flow. These numerical
calculations are carried out using a modified version of the 
{\sc zeus-MP} MHD code \citep{stone_hd,stone_mhd,hayes06}. In addition
to the standard ideal-MHD equations, we also solve for the ionization
fraction in the flow and solve the radiative transfer problem for
incoming ionizing photons (see \citealt{owenadams} for further
detail).

One constraint on this approach is that we assume that the
recombination time is short compared to the flow time and that the
mean-free path of the ionizing photons is short compared to the flow
length-scale at the ionization front. These assumptions are valid only
for the largest UV fluxes ($F_{UV}\gta~10^{5}$~erg~s$^{-1}$, values
appropriate for young Sun-like stars), so that the gas is close to
radiative-recombination equilibrium. In this regime, the thermal
structure in the flow is simplified, where ionized gas is nearly
isothermal at $T=10^{4}$~K and neutral gas is nearly isothermal at
$T=10^{3}$~K. Even with such high UV fluxes, the simulations show that
the outflows are safely controlled by the magnetic fields
\citep{owenadams}. Moreover, if the outflow is magnetically dominated
at high fluxes, it will also be magnetically dominated at lower fluxes
since the mass-loss rate increases with increasing flux.

The calculations are performed using a 2D spherical grid
(${r,\theta}$) with the assumption of azimuthal symmetry; the inner
boundary is set at $r=10^{10}$~cm and the outer boundary at
$r=1.5\times10^{11}$ cm ($R_P \le r \le 15 R_P$).  For comparison, the
outflow passes through the sonic point when the radius $r\approx3R_P$.
The radial grid is non-uniform with size $N_r=128$, where the
resolution at the inner boundary is sufficient to resolve the scale
height of the underlying atmosphere. In the angular direction we use a
uniform grid with 64 cells per quadrant. At the inner boundary we
apply fixed boundary conditions where the density
$\rho=10^{-11}$~g~cm$^{-3}$, the temperature $T=10^{3}$~K, the
magnetic field is a dipole with strength $B_P$, and the ionization
fraction $X=10^{-5}$. On the outer boundary we adopt outflow boundary
conditions, but include the contribution from the background stellar
field (controlled by the parameter $\beta$; see equation
[\ref{bfield}]). Finally, on the angular boundaries we adopt the
appropriate symmetry boundary conditions (see \citealt{owenadams} for
further detail).  In order to isolate the effects of the magnetic
field, we neglect the small contributions from planetary rotation and
the stellar gravitational field (see \citealt{trammell2014}). 

\begin{figure}
\plotone{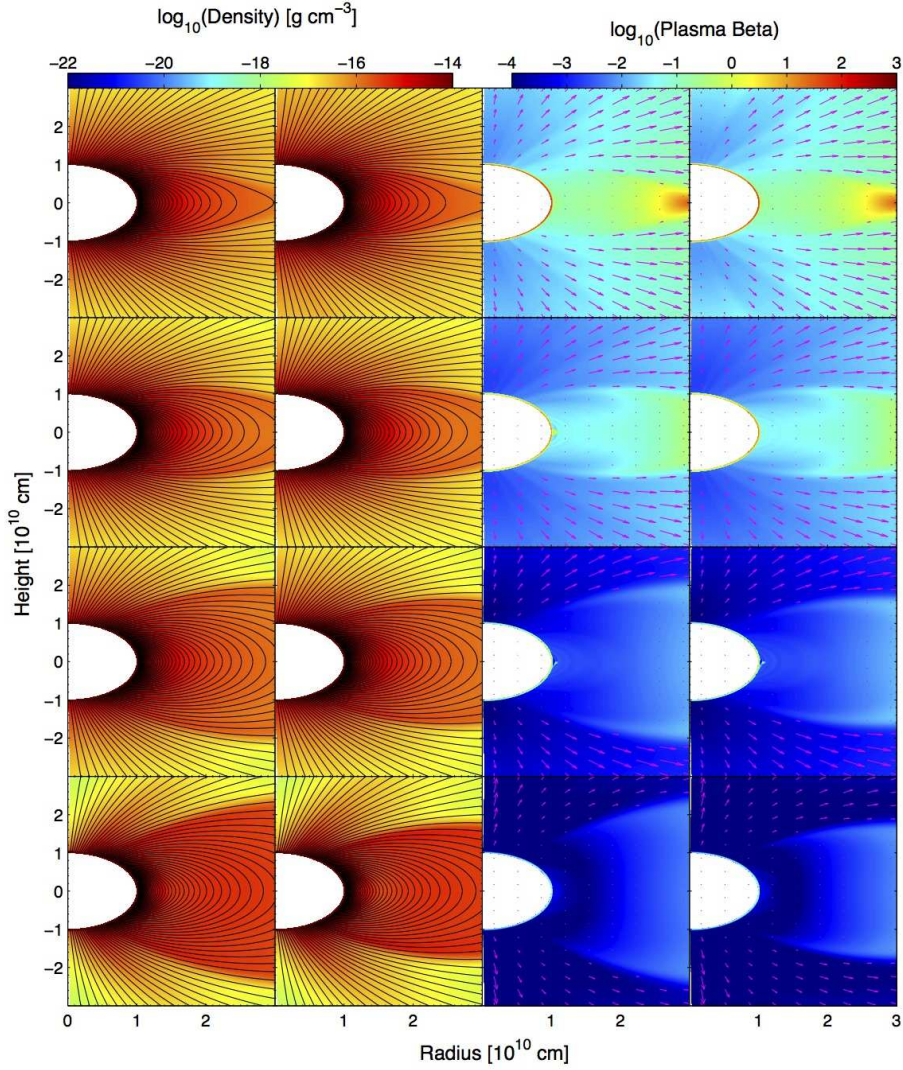}
\caption{Flow topologies for a subset of parameter space where both 
the stellar background field and the magnetic dipole of the planet
point in the $\hat z$ direction.  The rows represent planetary
magnetic field strengths from $B_P$ = 0.5 (top) to 10 (bottom)
gauss. The first two columns show the density and magnetic field
topology, where $\beta=0$ ($\beta=3\times10^{-3}$) for the first
(second) column. The final two columns show the plasma beta and
velocity structure, where $\beta=0$ ($\beta=3\times10^{-3}$) for the
third (fourth) column. The panels show only the inner regions of the 
simulations (which extend out to 15 planetary radii). The star is 
located along the positive $x$-axis.} 
\label{fig:one} 
\end{figure}  

Previous simulations have shown that the outflow is highly suppressed
from the night side of the planet \citep{owenadams}; as a result, this
work considers only the day side. This previous work carried out a
survey of the relevant parameter space, with variations in the field
strength ratio $\beta$, the planetary surface magnetic field strengths
$B_P$, and the UV flux $F_{UV}$ from the host star. All of these
simulations assume that the background stellar field is anti-aligned
with the dipole of the planet. Representative results from these
simulations are shown in Figure \ref{fig:one}, which presents the
outflow patterns and magnetic field configurations for planets
subjected to intense UV radiation fields with flux $F_{UV}$ = $10^{6}$
erg~s$^{-1}$~cm$^{-2}$. The figures shows the results obtained for
four planetary magnetic field strengths ($B_P$ = 0.5, 1, 4.0 and 10
gauss from top to bottom) and two values of the stellar magnetic field
parameter ($\beta$ = 0 and $3\times10^{-3}$ from left to right). The
first two columns show density and magnetic field topology; the second
two columns show the plasma beta and velocity structure. 

The results shown in Figure \ref{fig:one} show several trends. As the
planetary field strength $B_P$ decreases, the evaporative flow is able
to open up an increasing number of magnetic field lines and thereby
produce higher mass outflow rates. As the background field strength
parameter $\beta$ increases, a similar effects takes place as the
background field opens up more field lines and thereby allows mass
loss to take place from a larger fraction of the planetary surface
(see \citealt{adams2011} for an analytic treatment of this latter
effect). Notice that for relatively weak planetary fields, $B_P\lta1$
gauss, the opening of field lines is primarily due to the outflow
itself; for stronger fields, $B_P\gta1$ gauss, the number of opening
field lines depends strongly on the strength (and also topology) of
the background stellar field. We emphasize that these results were
obtained for large stellar UV fluxes. For smaller values of $F_{UV}$,
the background stellar field will dominate the process of field line
opening for smaller surface field strengths $B_P$. 

For all of the simulations with magnetic fields included, we find that
the mass outflow rates are smaller than those obtained in the absence
of fields. This suppression is significant, approximately an order of
magnitude, although the ratio varies with the other parameters of the
problem (see \citealt{owenadams}). Finally, we note that even for the
highest possible UV fluxes, which drive the most energetic outflows,
the flow is magnetically controlled. Although the numerical treatment
allows for the magnetic field lines to evolve in response to the fluid
motions, the magnetic field configurations do not change significantly
from their starting geometries over the course of the simulations. 

\begin{figure}
\plotone{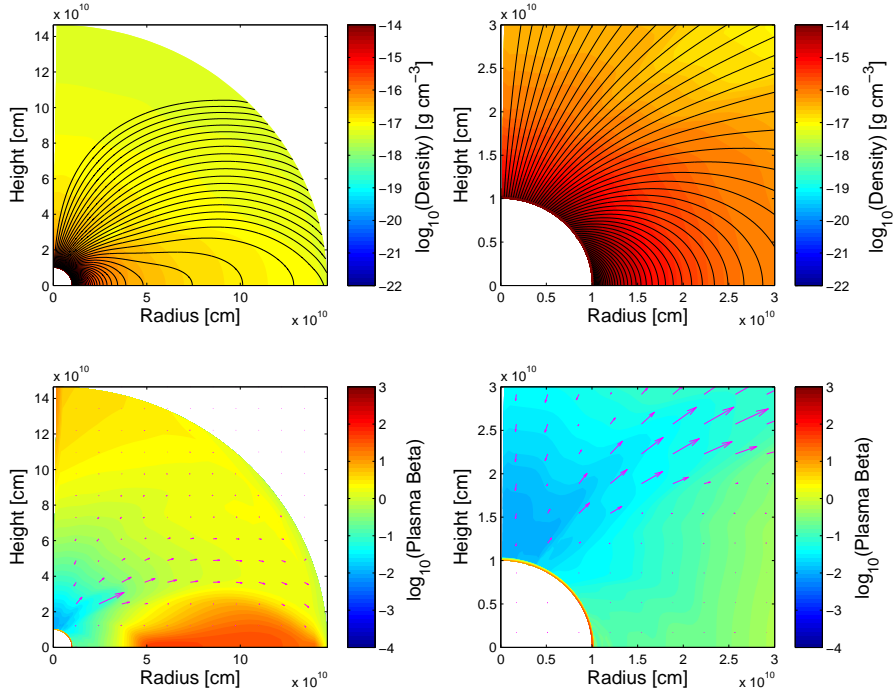}
\caption{Simulation of planetary outflow where the stellar background  
field has a split monopole configuration. Here the stellar field
points in the $\hat x$ direction (in the coordinate system centered on
the planet), whereas the magnetic dipole of the planet points in the
$\hat z$ direction. The planetary field $B_P$ = 0.5 gauss, the field
strength ratio $\beta = 3 \times 10^{-3}$, and the stellar UV flux
$F_{UV}=10^6$ erg cm$^{-2}$ s$^{-1}$. }
\label{fig:two} 
\end{figure} 

For comparison, Figure \ref{fig:two} shows the result from an
analogous numerical simulation, where the background stellar field has
a different direction (and is not aligned with the dipole of the
planet). The star is assumed to have a split-monopole configuration,
as expected when the stellar wind is strong enough to break open the
stellar field lines. From the small-scale viewpoint of the planet, the
background stellar field is uniform, but points in the $\hat x$
direction (radially towards the star). Results are shown for the
positive-$x$, positive-$z$ quadrant; since the mid-plane boundary
conditions assume that the star has a split monopole field, the
density structure of the flow has mid-plane reflection symmetry.
Although the magnetic fields are relatively weak (with $B_P$ = 0.5
gauss and $\beta = 3 \times 10^{-3}$), and the UV flux is relatively
strong (with $F_{UV}$ = $10^6$ erg cm$^{-2}$ s$^{-1}$), the magnetic
fields almost completely suppresses the outflow, i.e., flow takes place
only along a highly limited fraction of the field lines.

Comparison of Figures \ref{fig:one} and \ref{fig:two} shows that the
geometry of the background field (due to the star) can play an
important role in shaping planetary outflows. In this specific case,
the background geometry of the split-monopole field (shown in Figure
\ref{fig:two}) acts to suppress the mass loss rate (recall that even
aligned background fields suppress outflow rates relative to the
field-free case). Although not shown in Figure \ref{fig:two}, the flow
at high latitudes shows time-variability. This departure from
steady-state flow is observed in numerical simulations where the
outflow has difficulty passing through the sonic point
\citep{owenadams} and depends quite sensitively on the background
magnetic field \citep{adams2011}. 

\section{Conclusion} 

This contribution reviews recent progress concerning magnetically
controlled outflows from Hot Jupiters and presents results comparing
different configurations for the background magnetic field due to the
star. The following results emerge from this work: 

[1] Planetary outflows are magnetically controlled. Even in the
limiting case of relatively weak fields ($B_P \sim 0.5$ gauss) and
intense UV heating ($F_{UV} = 10^6$ erg cm$^{-2}$ s$^{-1}$) the 
magnetic field lines guide the flow.

[2] Magnetic fields act to suppress the mass outflow rates relative to
the values obtained in the $B\to0$ limit (and relative to the
energy-limited expression of equation [\ref{estimate}]). As a rule, 
the outflow rates are suppressed by an order of magnitude. 

[3] Magnetic fields change the geometry of the flow. Not all field
lines are open and can support outflow. The fraction of open field
lines is determined mostly by the background magnetic field structure,
but also by thermal pressure opening up planetary field lines. In
addition, magnetic fields suppress the transfer of heat to the night
side of the planet and suppress outflow from that hemisphere.

[4] The manner in which the magnetic fields of the planet match onto
the background environment, provided by the stellar magnetic field and
the stellar wind, is important. The geometry of the background
magnetic field affects the fraction of the planetary surface that
supports open field lines and hence outflow (compare Figures
\ref{fig:one} and \ref{fig:two}).

\acknowledgments{We would like to thank the organizers of Cool Stars  
18 for inviting this contribution.  The numerical calculations were
performed on the Sunnyvale cluster at CITA, which is funded by the
Canada Foundation for Innovation. We are grateful for the hospitality
of both CITA and the University of Michigan for visits that helped
facilitate this collaboration.}

\end{document}